\documentclass[showpacs,aps, twocolumn]{revtex4-1}
\usepackage{epsfig}
\usepackage{graphicx}
\usepackage{amsmath,amssymb,amsfonts}
\usepackage{array}
\usepackage{url}
\usepackage{hyperref}
\usepackage{multirow}
\usepackage{float}
\usepackage{lineno}
\usepackage{xspace}
\usepackage{ulem}
\usepackage{lipsum} 
\usepackage[usenames,dvipsnames]{color}
\begin{document}
\title{Global polarization of $\Lambda$ hyperons in hot QCD matter at TeV energies}
\author{Bhagyarathi Sahoo}
\email{Bhagyarathi.Sahoo@cern.ch}
\author{Captain R. Singh}
\email{captainriturajsingh@gmail.com}
\author{Raghunath Sahoo} 
\email{Corresponding Author: Raghunath.Sahoo@cern.ch}

\affiliation{Department of Physics, Indian Institute of Technology Indore, Simrol, Indore 453552, India}

\begin{abstract}
The study of spin polarization of $\Lambda$ hyperons in ultrarelativistic heavy-ion collisions provides insights into the angular momentum and vortical structure of the possible existence of QGP. The present study examines the global spin polarization of $\Lambda$ hyperons using a second-order relativistic viscous hydrodynamic framework that incorporates medium vorticity, shear viscosity, and evolving magnetic fields. It explores thermal vorticity evolution in relativistic heavy-ion collisions and evaluates its value at the decoupling isothermal freeze-out surface. We quantify the contributions of thermal vorticity and magnetic field to the global spin polarization of $\Lambda$ hyperons. Comparing results with recent ALICE measurements in Pb+Pb collisions at $\sqrt{s_{NN}}$ = 2.76 and 5.02 TeV shows qualitative agreement, offering new insights into the vortical structure of QCD matter. It also explores the relationship between magnetic and rotational dynamics, with implications for spin polarization at RHIC and LHC energies.

\end{abstract}
\date{\today}
\maketitle

\section{Introduction}
\label{intro}

Over the last few decades, considerable efforts have been made to study the role of orbital angular momentum (OAM) and its implications in relativistic heavy-ion collisions~\cite{Liang:2004ph, Liang:2004xn, Becattini:2007sr, STAR:2017ckg}. Theoretical predictions, supported by experimental observations, suggest that peripheral heavy-ion collisions may generate an initial OAM of the order of $10^{5}\rm\hbar$~\cite{Becattini:2007sr, STAR:2017ckg}. A substantial fraction of 
initial OAM is thought to be deposited in the deconfined phase of QCD matter called quark–gluon plasma (QGP), where it manifests as rotational motion or vorticity within the system. This vorticity can lead to the spin polarization of hadrons through spin-orbit coupling. In this context, the global spin polarization of hyperons and the spin alignment of vector mesons observed in relativistic heavy-ion collisions are considered as direct manifestations of the vortical dynamics of the QGP. Following the landmark measurement of $\Lambda$ hyperons spin polarization by the STAR experiment~\cite{STAR:2017ckg}, a broad range of theoretical and phenomenological studies have further advanced this field~\cite{Karpenko:2016jyx, Alzhrani:2022dpi, Vitiuk:2019rfv, Sun:2017xhx, Ivanov:2019ern, Ivanov:2022ble, Wei:2018zfb, Becattini:2024uha, Guo:2021udq, Ryu:2021lnx}. Hydrodynamic and transport model calculations of the global spin polarization of $\Lambda$ hyperons exhibit qualitative agreement with measurements at Relativistic Heavy Ion Collider (RHIC) and HADES energies~\cite{Alzhrani:2022dpi, Vitiuk:2019rfv, Sun:2017xhx, Ivanov:2019ern, Ivanov:2022ble, Wei:2018zfb, Becattini:2024uha, Guo:2021udq, Ryu:2021lnx}. These frameworks employ a thermally equilibrated treatment of spin degrees of freedom in describing polarization phenomena. In addition to $\Lambda$ hyperons, the averaged global spin polarization measurement of $\Omega$ and $\Xi$ hyperons is obtained using the polarization transfer method in the STAR  collaboration~\cite{STAR:2020xbm}. Collectively, these studies characterize medium as the most vortical fluid with the least viscosity produced in relativistic heavy-ion collisions under extreme temperature and density.\\

In addition to global spin polarization~\cite{STAR:2018gyt, STAR:2021beb, STAR:2023nvo}, detailed investigations have established the local (longitudinal)~\cite{STAR:2019erd, Becattini:2017gcx, Becattini:2021iol, ALICE:2021pzu, STAR:2023eck, Sahoo:2024egx}, transverse~\cite{Qin:2025cvp, STAR:2025jwc, LHCb:2025rxf}, and helicity~\cite{Gao:2021rom, Yi:2021unq, Yi:2023tgg} spin polarization of $\Lambda$ hyperons across RHIC and LHC energies. Similarly, for spin-1 vector mesons, global spin alignment measurements of $\phi$, $K^{*0}$, $D^{*+}$, J/$\psi$, and $\Upsilon(1S)$ have been carried out over a wide range of collision systems at RHIC and LHC facilities~\cite{STAR:2022fan, STAR:2022fan, STAR:2008lcm, ALICE:2019aid, ALICE:2023jad, ALICE:2020iev, STAR:2013iae, STAR:2020igu, ALICE:2025cdf, Sahoo:2023oid, Sahoo:2025kur, Sahoo:2025bkx}.
In theoretical studies, multiple sources contributing to spin polarization have been identified. Alongside the vorticity field, electromagnetic fields, correlations and fluctuations of strong interaction fields (such as vector meson field), shear viscosity, gradients of temperature and baryon chemical potential, medium anisotropy, as well as chromo-electric and chromo-magnetic fields, play important roles in shaping the spin polarization of hyperons~\cite{Csernai:2018yok, Sahoo:2024yud, Yang:2017sdk, Fu:2021pok} and the spin alignment of vector mesons~\cite{Sheng:2022wsy, Xia:2020tyd, Mohanty:2021vbt}. Motivated by this comprehensive set of contributing mechanisms, a quantitative assessment of their individual roles requires a unified, systematic framework. It leads to the formulation of a modern hybrid relativistic viscous spin-magnetohydrodynamic approach, where hydrodynamic fields and their leading and higher-order gradients are consistently incorporated, and spin is treated as an explicit dynamical degree of freedom for the evolution of the medium.\\

To systematically quantify and disentangle the various contributions to spin polarization, we developed a hydrodynamic framework that consistently incorporates medium vorticity, viscosity, and magnetic fields within a M\"{u}ller–Israel–Stewart (MIS) based second-order relativistic viscous hydrodynamics. Within this framework, we evaluate the global spin polarization of $\Lambda$ hyperons using hydrodynamic initial conditions tuned to describe Pb–Pb collisions at $\sqrt{s_{\rm NN}} = 2.76$ and $5.02$ TeV. The space-time evolution of the medium is obtained by solving the coupled rate equations, thereby enabling estimates of the vortical QGP lifetime across different scenarios.
Furthermore, the evolution of thermal vorticity is computed using a vorticity dissipation rate equation that incorporates the effects of viscosity, constant magnetic fields, time-dependent magnetic fields, and scenarios without magnetic fields. The resulting thermal vorticity on the isothermal decoupling freeze-out hypersurface is then used to determine the spin polarization of $\Lambda$ hyperons in two representative cases: one with a time-dependent magnetic field and another without it. The relation between thermal vorticity and the mean spin vector at global equilibrium is utilized to extract the spin polarization observable. Further,  results are compared with recent ALICE measurements of global $\Lambda$ hyperons spin polarization in Pb–Pb collisions at $\sqrt{s_{\rm NN}} = 2.76$ and $5.02$ TeV for (15–50)\% centrality. We further analyze the respective roles of thermal vorticity and magnetic fields in shaping the observed polarization. This study offers a coherent understanding of spin polarization phenomena across a broad range of collision energies, spanning experiments from HADES to RHIC and the LHC. \\

This paper is organized as follows. Following the introduction in section~\ref{intro}, we present the details of the hydrodynamic framework employed in this study in section~\ref{formulation}. The results obtained from the model are described in section~\ref{res}. Finally, section~\ref{sum} summarizes the main findings and outlines possible directions for further refinement of the framework.

\section{Hydrodynamic Setup}
\label{formulation}
This section provides a detailed formulation of the extended relativistic hydrodynamic framework used in our study. Special emphasis is placed on implementing the effects of rotation and magnetic fields within the causal, second-order M\"{u}ller-Israel–Stewart (MIS) formulation. This approach results in coupled evolution equations for temperature, viscosity, and vorticity. Additionally, we establish a formalism for computing the spin polarization of $\Lambda$ hyperons based on this framework.\\

The total energy-momentum tensor of the rotating, viscous, and magnetized fluid is defined as~\cite{Denicol:2018rbw, Roy:2015kma, Biswas:2020rps, Pu:2016ayh};

\begin{equation}
\label{TmunuB}
T^{\mu\nu} =  \left( \epsilon+P+B^{2}\right) u^{\mu}u^{\nu} -g^{\mu\nu}\left(P+\frac{B^{2}}{2}\right) - B^{\mu}B^{\nu}  + \pi^{\mu\nu}
\end{equation}

\noindent
here, $\epsilon $ is energy density, $P$ is pressure, $B^{\mu} = \frac{1}{2}\epsilon^{\mu\nu\alpha\beta}F_{\nu\alpha}u_{\beta}$ is the magnetic field in the frame moving with the velocity $u_{\beta}$, and $F_{\nu\alpha}$ is the field strength tensor. Here, $\epsilon^{\mu\nu\alpha\beta}$ is the Levi-Civita antisymmetric tensor. The magnetic field four vector $B^{\mu}$ is  space-like four vector with modulus $B^{\mu}B_{\mu} = -B^{2}$ and orthogonal to $u^\mu$, i.e., $B^{\mu}u_{\mu} = 0$, where $B = |\vec{\textbf{B}}|$, and  $|\vec{\textbf{B}}|$ is the magnetic three vector. 
Further, we have considered the four magnetic vector $B_{\mu} = (0, 0, B_y, 0)$, and the velocity profile  $ u^{\mu} =\gamma(1, v_{x}, 0, v_{z} )$, where $\gamma = \frac{1}{\sqrt{1-\vec{v}^{2}}}$ being the Lorentz factor.\\ 

In Eq.~\ref{TmunuB}$, \pi^{\mu\nu}$ is the shear stress tensor and defined as,
\begin{equation}
\label{eq8}
\pi^{\mu\nu} = \eta_{s} \bigtriangledown^{<\mu}u^{\nu>}
\end{equation}
where $\eta_{s}$ is the shear viscosity, with:
\begin{equation}
\bigtriangledown^{<\mu}u^{\nu>} \equiv 2\bigtriangledown^{(\mu}u^{\nu)} - \frac{2}{3}\Delta^{\mu\nu}\bigtriangledown^{\alpha} u_{\alpha} 
\nonumber
\end{equation} 
the $\bigtriangledown^{(\mu}u^{\nu)}$ is described as $A^{(\mu}B^{\nu)} = \frac{1}{2} \left(A^{\mu}B^{\nu} + A^{\nu}B^{\mu} \right)$. \\

\noindent
The energy density of a viscous medium in the presence of a magnetic field can be obtained by solving the energy-momentum conservation ($\partial_{\mu}{T^{\mu\nu}} = 0$)~\cite{Pu:2016ayh, Roy:2015kma, Biswas:2020rps, Sahoo:2023xnu, Muronga:2001zk, Muronga:2003ta};

\begin{equation}
\label{DepsilonB}
D \epsilon = -\left( \epsilon+P+B^2 - \Phi \right)\;\theta - B D B
\end{equation}
where, $D \equiv u^{\mu}\partial_{\mu}$ is the convective time derivative and $\theta\equiv \partial_{\mu}u^{\mu}$ is the volume expansion.The viscous term 
$\Phi = \pi^{00} - \pi^{zz}$ is the difference between temporal and spatial components of the shear viscosity tensor representing the viscous term.
The second-order MIS relaxation equation using Grad's  14 moments methods for shear viscosity has the following form \cite{Sahoo:2023xnu, Muronga:2001zk, Muronga:2003ta};

\begin{align}
D\pi^{\mu\nu} &= - \frac{1}{\tau_{\pi}}\pi^{\mu\nu}- \frac{1}{2\beta_{2}}\pi^{\mu\nu}\left[\beta_{2}\theta + TD\left(\frac{\beta_{2}}{T}\right)\right] \nonumber \\&+ \frac{1}{\beta_{2}} \bigtriangledown^{<\mu}u^{\nu>},
\label{phievolution}
\end{align}
where,
$\tau_{\pi} = 2\eta_s\beta_{2}$ is the relaxation time, $\beta_{2}$ is the relaxation coefficient given as; $\beta_{2} = 3/4P$.  
Here shear viscosity $\eta_s = bT^{3}$. 
Pressure profile of the medium is $P = \epsilon/3= aT^{4} $ at zero baryon chemical potential. The parameter $a$ and $b$ in Eq.~\ref{phievolution}, are defined as~\cite{Sahoo:2023xnu, Muronga:2001zk, Muronga:2003ta};

\begin{equation}
a = \frac{\pi^{2}}{90} \left[ 16 + \frac{21}{2}N_{f} \right],
\label{a}
\end{equation}
\begin{equation}
b = (1 + 1.70 N_{f})\frac{0.342}{(1 + N_{f}/6)\alpha_{s}^{2}\ln(\alpha_{s}^{-1})} 
\label{b}
\end{equation}
where $N_{f} = 3$, is the number of flavor and $\alpha_s$ is the running coupling constant.\\

The rotational dynamics of the medium strongly depend on the choice of velocity profile. Therefore, the velocity profile is chosen in such a way that the transverse component of velocity ($v_{x}$) depends on the 
longitudinal component and the longitudinal component of velocity ($v_{z}$) on the transverse component~\cite{Sahoo:2023xnu, Singh:2018bih}. Velocity along the rotation axis is considered to be zero, i.e., $v_{y} = 0$. The ($v_{x}$) and ($v_{z}$) are given below;

\begin{equation}
\label{vxz}
v_{x} =  \frac{\omega z}{2}, \;\;\;\;
v_{z} =  \frac{z}{\tau} - \frac{\omega x}{2}
\end{equation}
In Eq.~\ref{vxz}, $\omega$ denotes the vorticity of the system, and plays a central role in determining the spin polarization tensor $\omega_{\mu\nu}$. The spin polarization tensor is obtained using a tensor decomposition, and given as~\cite{Florkowski:2017ruc}; 

\begin{equation}
\omega_{\mu\nu} = k_{\mu}u_{\nu} - k_{\nu}u_{\mu} + \epsilon_{\mu\nu\alpha\beta}u^{\alpha}\omega^{\beta}
\label{spinpolarization}
\end{equation}
where, $k_{\mu}$ and $\omega_{\mu}$ are defined in terms of spin polarization tensor;
\begin{equation}
k_{\mu} = \omega_{\mu\nu}u^{\nu}, \;\;\;\;\;\;\;\;\; \omega_{\mu} = \frac{1}{2} \epsilon_{\mu\nu\alpha\beta}\omega^{\nu\alpha}u^{\beta}
\label{momentumvector}
\end{equation}

The orthogonality conditions $k^{\mu}u_{\mu} = \omega^{\mu}u_{\mu} = 0$ are imposed by construction, ensuring that both $k_{\mu}$ and $\omega_{\mu}$ remain normal to the fluid four-velocity $u_{\mu}$.  
Focusing on a configuration with rotation in the $x$–$z$ plane, specified by $\omega_{\mu} = (0,0,\omega,0)$, Eqs.~\ref{spinpolarization} and \ref{momentumvector} are solved self-consistently to determine the spin polarization tensor $\omega_{\mu\nu}$.

\begin{equation}
\omega_{\mu\nu} =
\left[ {\begin{array}{cccc}
0 & 0 & 0 & 0 \\
0 & 0 & 0 & \frac{\omega}{T} \\
0 & 0 & 0 & 0 \\
0 & -\frac{\omega}{T} & 0 & 0 \\
\end{array} } \right]
\label{polarizationmatrix}
\end{equation}

\noindent
It is worthwhile to mention that in both global and local equilibrium scenarios, the spin polarization tensor $\omega_{\mu\nu}$ coincides with the thermal vorticity $\bar{\omega}_{\mu\nu}$, i.e., $\omega_{\mu\nu} = \bar{\omega}_{\mu\nu}$. The distinction lies in the behavior of $\beta^{\mu} = \frac{u^{\mu}}{T}$, which serves as a Killing vector in global equilibrium, whereas in local equilibrium, it follows a distinctive space-time dependence~\cite{Florkowski:2018ahw}. \\

Now, to quantify the effect of rotation on the thermal evolution of the medium, vorticity has been incorporated in Euler's thermodynamic relation, written as;

\begin{equation}
\epsilon + P = Ts + \mu n + \Omega \rm w + eBM 
\label{eulerequation}
\end{equation}
where, $\Omega$ denotes the chemical potential associated with rotation, while $\rm w$ represents the rotation density. These variables are defined as $\Omega = \frac{T}{2\sqrt{2}}\sqrt{\omega_{\mu\nu}\omega^{\mu\nu}} $ and $\rm w = 4n_{0}cosh(\xi)$, where $\xi = \frac{\omega}{2T}$ and $n_{0} = \frac{T^{3}}{\pi^{2}}$ corresponds to the particle number density under the massless limit. Accordingly, the rotational density takes the form $\rm w = 4\frac{T^{3}}{\pi^{2}}\cosh\left(\frac{\omega}{2T}\right)$~\cite{Singh:2018bih}. In addition, the magnetization of the fluid is expressed as $M = \chi_{m} B$, where $\chi_{m}$ denotes the magnetic susceptibility and $B$ stands for the magnetic field. In this study, a time-dependent magnetic field is considered with an exponential decay profile given by $B(\tau) = B_{0} \exp(-\tau/\tau_{B})$. Here, $\tau_{B}$ denotes the magnetic decay parameter, characterizing the timescale over which the magnetic field persists.\\

Thus, taking all the above inputs at zero baryonic chemical potential, Eq.~\ref{eulerequation} can be modified as,
\begin{equation}
\epsilon + P = Ts + \frac{2\omega T^{3}}{\pi^{2}} \cosh\left(\frac{\omega}{2T}\right) + e\chi_{m} B^{2}
\label{eulerequation1}
\end{equation}

Substituting Eq.~\ref{eulerequation1} in Eq.~\ref{DepsilonB}, we have
\begin{align}
D \epsilon & =  - \left[ Ts + \frac{2\omega T^{3}}{\pi^{2}} \cosh\left(\frac{\omega}{2T}\right) + (1+e\chi_{m})B^{2} - \Phi \right]\; \theta \nonumber \\ & - B D B
\label{eulerequation3}
\end{align}

Moreover, under the ideal limit, $\epsilon = 3P$, the above Eq.~\ref{eulerequation1} becomes,
\begin{equation}
\label{energy}
\epsilon = \frac{3}{4}\bigg[ Ts + \frac{2\omega T^{3}}{\pi^{2}} \cosh\left(\frac{\omega}{2T}\right) + e\chi_{m} B^{2} \bigg]
\end{equation}
Now, taking the derivative of Eq.~\ref{energy} w.r.t. $D$, and using the entropy relation $s = c+dT^{3}$ (c and d are the constants),  we get;

\begin{equation}
\label{dedtau2}
D\epsilon = \frac{3}{4}\bigg[ \bigg(s+3dT^{3}+\frac{2F}{\pi^{2}}\bigg)DT + \frac{2G}{\pi^{2}} D\omega + 2e\chi_{m} B D B \bigg], 
\end{equation}
where, $F= 3T^{2}\omega \cosh\left(\frac{\omega}{2T}\right) - \frac{1}{2} \omega^{2}T \sinh\left(\frac{\omega}{2T}\right)$ and
$G = T^{3} \cosh\left(\frac{\omega}{2T}\right) + \frac{1}{2} \omega T^{2}  \sinh\left(\frac{\omega}{2T}\right)$. \\

Comparing Eq. \ref{dedtau2} and Eq. \ref{eulerequation3}, one can obtain the vorticity evolution equation, given as;

\begin{align}
\label{wevolution}
D\omega = \frac{-\pi^{2}}{2G} \bigg[ \frac{4}{3}\bigg(\epsilon+ P+ B^2  - \Phi \bigg)\; \theta & + \bigg(\frac{4}{3} + 2e \chi_m \bigg) B D B \nonumber\\  + \bigg(s+3dT^{3}+\frac{2F}{\pi^{2}}\bigg)DT\bigg]
\end{align}

Following this, the formulation leads to three non-linear, coupled differential equations, Eqs.~\ref{eulerequation3}, \ref{phievolution}, and \ref{wevolution}, which govern the evolution of the medium in terms of energy density (or temperature), viscous effects, and vorticity, respectively. The explicit forms of the operators $D$ and $\theta$ are provided in Appendices~\ref{D} and \ref{theta}. The complete expressions of these coupled equations in Milne coordinates, corresponding to the specified velocity profile, are derived in Appendices~\ref{tempcooling}, \ref{shearviscous}, and \ref{vordiss}.
For $\omega = 0$, the formulation reduces to second-order viscous hydrodynamics, while in the additional limit $\Phi = 0$, it recovers the ideal fluid description.\\

\noindent
Employing the vorticity and temperature profiles evaluated at the freeze-out hypersurface, the spin polarization of hyperons is determined. At leading order, the mean spin vector of a spin-$1/2$ particle with four-momentum $p_{\delta}$, arising from thermal vorticity and electromagnetic fields, is given by~\cite{Becattini:2016gvu, Guo:2019joy},

\begin{equation}
S^{\mu}(x, p) = -\frac{1}{8m} (1 - f(x,p)) \epsilon^{\mu \rho \sigma \delta}p_{\delta} \bigg[\bar{\omega}_{\rho \sigma} - \frac{2 \mu_{\Lambda}  F_{\rho \sigma}}{T} \bigg] 
\label{spinpolarization1}
\end{equation}

where m is the mass of the particle, $\mu_{\Lambda}$ is the magnetic moment for $\Lambda$ hyperons, $\mu_{\Lambda} \approx -0.61\mu_{N} \approx -\frac{0.61e}{2m_p}$. The $\bar{\omega}_{\rho \sigma}$ is the thermal vorticity and is defined as the anti-symmetric derivative of the four-temperature field,

\begin{equation}
\bar{\omega}_{\mu \nu} = -\frac{1}{2}(\partial_{\mu}\beta_{\nu}-\partial_{\nu}\beta_{\mu})
\label{thermalvorticity}
\end{equation}
\noindent
Here, $\beta^{\mu} = \frac{u^{\mu}}{T}$ and  $f(x,p)$ is the covariant Fermi-Dirac phase space distribution,
given by,

\begin{equation}
f(x,p) = \frac{1}{\exp{\left[\beta. p - \sum_{j}{\mu_j q_j/T}\right]} +1}
\end{equation}

Now, using the co-moving axial thermal vorticity and co-moving magnetic field, the spin polarization vector can be written as,

\begin{equation}
S^{\mu}(x, p) = \frac{1}{4} (1 - f(x,p)) \bigg[\bar{\omega}^{\mu} - \frac{2 \mu_{\Lambda}  B^{\mu}}{T} \bigg] 
\label{spinpolarization1}
\end{equation}
\noindent
Below, we have obtained  the mean spin vector by  integrating Eq.~\ref{spinpolarization1} over the decoupling freeze-out hypersurface $\Sigma$,
\begin{equation}
S^{\mu}(p) = \frac{\int_{\Sigma} d\Sigma_{\lambda}p^{\lambda} \hspace{0.1cm} f(x,p) S^{\mu}(x, p)}{\int_{\Sigma} d\Sigma_{\lambda}p^{\lambda} \hspace{0.1cm}f(x,p)},
\label{spinpolarization2}
\end{equation}

Equation~\ref{spinpolarization2} can be further expanded as follows:
\begin{equation}
S^{\mu}(p) = \frac{1}{4} \frac{\int_{\Sigma} d\Sigma_{\lambda}p^{\lambda} \hspace{0.1cm} f(x,p)(1-f(x,p)) \bigg[\bar{\omega}^{\mu} - \frac{2 \mu_{\Lambda}  B^{\mu}}{T} \bigg]}{\int_{\Sigma} d\Sigma_{\lambda}p^{\lambda} \hspace{0.1cm}f(x,p)},
\label{spinpolarization3}
\end{equation}
\noindent

The integrals of Eq.~\ref{spinpolarization3} can be solved using the cylindrical coordinates and can be written as
\begin{equation}
S_y  = \frac{1}{4 \; T_{\rm dec}} \frac{\displaystyle\int   \tau r\; \displaystyle\sum_{n=1}^{\infty} \; (\omega_{y} - 2 \mu_{\Lambda} B_{y})  m_T I_0 \left( O_{1}\right)  K_1 \left( O_{2}\right) dr }{\displaystyle\int \tau r  \displaystyle\sum_{n=1}^{\infty}  m_T I_0 \left(O_1 \right)  K_1 \left(O_2 \right) dr}
\label{spinpolarization7}
\end{equation}

\noindent
where, I's, K's are Bessel functions and $O_{1} = n \frac{p_T}{T} \sinh y_T$, $O_{2} = n \frac{m_T}{T} \cosh y_T$. Here, we have assumed that hadronization emerges on the decoupling hypersurface with a constant proper time $\tau$ and a temperature $T = T_{\rm dec}$. A detailed derivation of Eq.~\ref{spinpolarization7} is provided in Appendix~\ref{hypersurface}. In the rest frame of the particle, the spin vector is $ S^{*\mu} = (0, \bf{S^{*}})$, which is obtained by using the Lorentz transformation, given as:

\begin{equation}
\mathbf{S}^{*} = \mathbf{S(p)} - \frac{{\mathbf{S \cdot p}}}{E(E+m)}\bf{p}
\label{spinpolarization4}
\end{equation}
Finally, the net spin polarization is defined as;

\begin{equation}
    \textbf{P}  = 2 \textbf{S{*}}
\end{equation}

\section{Results and Discussion}
\label{res}

\begin{figure*}[ht!]
\centering
\includegraphics[scale = 0.5]{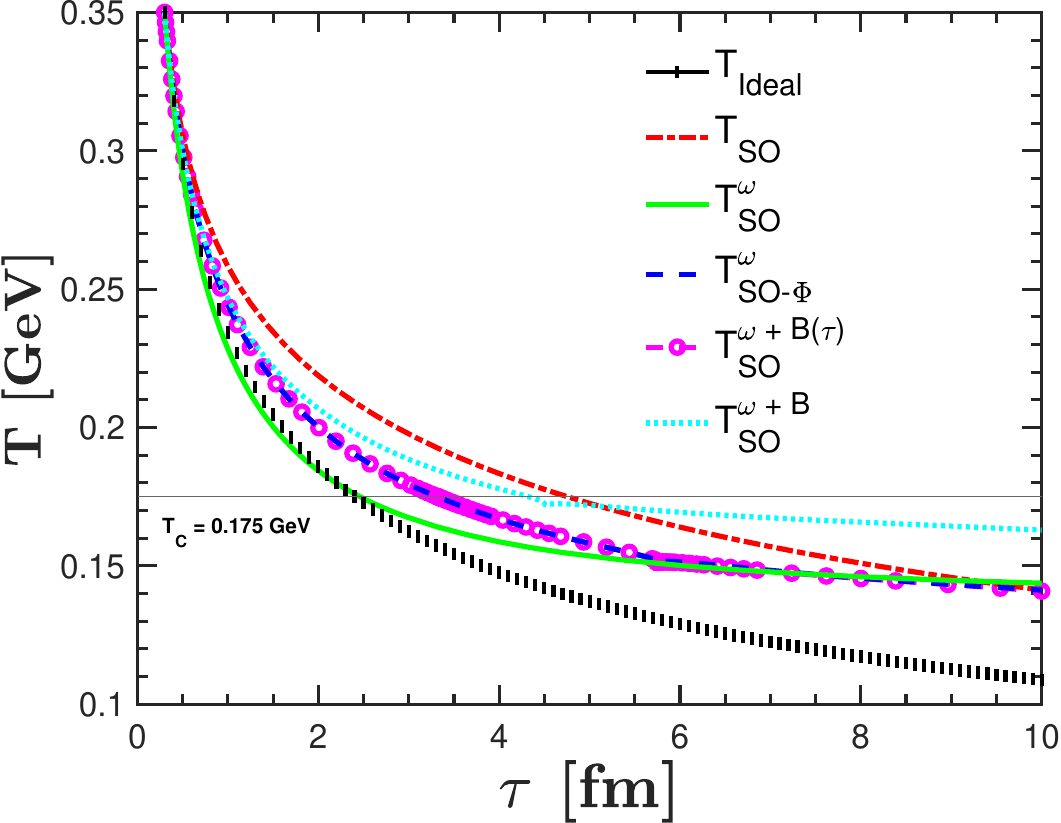}
\includegraphics[scale = 0.5]{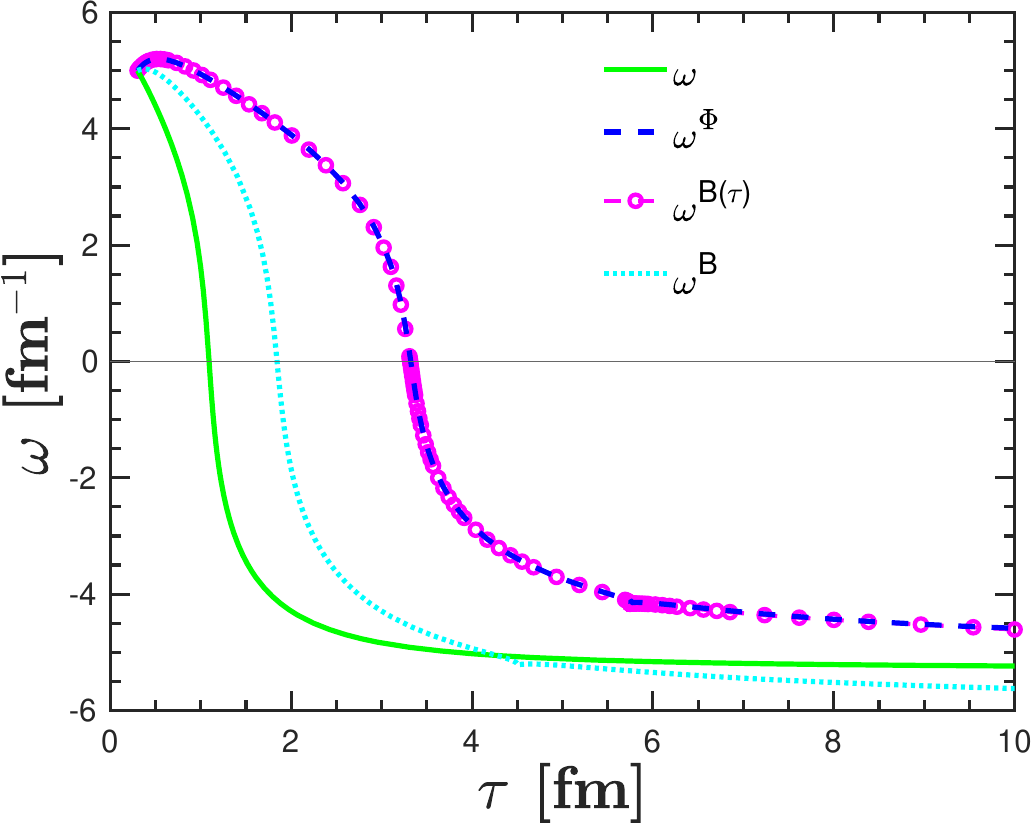}
\caption{(Color Online) The temperature (T), and vorticity ($\omega$) are plotted as a function of time $\tau$ in the left and right panel, respectively, with the initial conditions: {\bf T = 0.35 GeV, $\tau_{0}$ = 0.3 fm, $\omega_{0}$ = 5.0 fm$^{-1}$, $\Phi_{0}$ = 0.053 GeV$^4$}. In temperature plot,  For T$_{\text{Ideal}}$; $\omega = 0$, $\Phi = 0$, $B = 0$. For T$_{\text{SO}}$; $\omega = 0$, $\Phi \ne 0$, $B = 0$. For T$^{\omega}_{SO}$; $\omega \ne 0$, $\Phi = 0$, $B = 0$. For T$^{\omega}_{SO-\Phi}$; $\omega \ne 0$, $\Phi \ne 0$, $B = 0$. For T$^{\omega + B(\tau)}_{SO}$; $\omega \ne 0$, $\Phi \ne 0$, and magnetic field is changing with time, B($\tau$). For T$^{\omega + B}_{SO}$; $\omega \ne 0$, $\Phi \ne 0$, and the magnetic field is constant. The $\omega$ plot has the same meaning of notations as the temperature plot.} 
\label{fig1}
\end{figure*}

The hydrodynamic evolution of the medium is analyzed in the presence of dissipative effects, including viscosity, vorticity, and magnetic fields. Two representative magnetic field configurations are considered: a time-dependent field $B(\tau)$ and a constant field $B$. The individual and combined effects of these dissipative mechanisms on the evolution of temperature and vorticity are systematically examined.
The evolution of temperature, viscosity, and vorticity is governed by three non-linear coupled differential equations, whose explicit forms are provided in Appendices~\ref{tempevolution}, \ref{phievolution2}, and \ref{wevolution2eB}. These equations are solved at mid-rapidity using specified initial conditions: $T_{0}$, $\tau_{0}$, $\omega_{0}$, and $\Phi_{0}$. In the present study, initial thermalization time and temperature are taken as $\tau_{0} = 0.3$ fm and $T_{0} = 0.350$ GeV, respectively. The initial viscous term is given by $\Phi_{0} = \frac{1}{3\pi}\frac{s_{0}}{\tau_{0}}$ at $\tau = \tau_{0}$, where $s_{0}$ denotes the initial entropy density~\cite{Sahoo:2023xnu}. The initial vorticity $\omega_{0}$ is chosen to remain consistent with relativistic constraints (causality) on rotational motion; in this work, $\omega_{0} = 5\ \mathrm{fm}^{-1}$ is taken. The selected set of initial conditions is calibrated for Pb+Pb collisions in the (15–50)\% centrality class. Consistent with earlier studies~\cite{Sahoo:2023xnu}, the sensitivity of the hydrodynamic evolution to the choice of initial conditions is established, with each set defining a distinct dynamical evolution of the system. With the chosen initial conditions, subsequently, corresponding values of temperature and thermal vorticity at the isothermal freeze-out hypersurface are then used to determine the spin polarization of $\Lambda$ hyperons. \\

Figure~\ref{fig1} presents the temperature cooling (left panel) and vorticity dissipation (right panel) as functions of proper time $\tau$, obtained by solving the non-linear coupled differential equations for $T$, $\omega$, and $\Phi$ with the specified initial conditions. The left panel illustrates the temperature evolution under different medium scenarios, including the ideal limit (T$_{\text{Ideal}}$ corresponds to $\omega = 0$ and $\Phi = 0$), viscous-only (T$_{\text{SO}}$), vortical-only $\left( T^{\omega}_{\text{SO}} \right)$, and their combined effects, as well as cases with time-dependent and constant magnetic fields.
Further, notation for coupled viscosity and vorticity effects is denoted as $\left( T^{\omega}_{\text{SO}-\Phi} \right)$ with an additional time-dependent magnetic field $\left(T^{\omega + B(\tau)}_{\text{SO}-\Phi} \right)$ and with a constant magnetic field $\left( T^{\omega + B}_{\text{SO}-\Phi} \right)$. Notably, $T_{SO}$ $\left(T^{\omega}_{SO} \right)$ stands for the temperature obtained by solving the second-order hydrodynamic equations Eq.~(\ref{tempevolution}) and ~\ref{phievolution2} (\ref{wevolution2eB}) at $\omega=0$ ($\omega\neq 0$). The $T^{\omega}_{\text{SO}-\Phi}$ stands for the temperature obtained by solving the second-order hydrodynamic equations Eq.~(\ref{tempevolution}), ~\ref{phievolution2}, and (\ref{wevolution2eB}) in the absence of a magnetic field ($B = 0$). While $T^{\omega + B(\tau)}_{\text{SO}-\Phi}$ stands for the temperature obtained by solving the second-order hydrodynamic equations Eq.~(\ref{tempevolution}), ~\ref{phievolution2}, and (\ref{wevolution2eB}) with a time-varying magnetic field with profile $B(\tau) = B_{0} \exp({-\tau/\tau_{B}})$. At last,
$T^{\omega + B}_{\text{SO}-\Phi}$ stands for the temperature obtained by solving the second-order hydrodynamic equations Eq.~(\ref{tempevolution}), ~\ref{phievolution2}, and (\ref{wevolution2eB}) with a constant magnetic field. It is noteworthy to mention that for irrotational fluid ($\omega = 0$), the longitudinal boost invariant velocity profile is assumed, and the evolution is similar to a Bjorken-like flow. However, for the rotational fluid ($\omega \neq 0$), we consider the velocity profiles mentioned in Eq.~(\ref{vxz}).\\

In the absence of the dissipation effects, T$_{\text{Ideal}}$ reflects the fast cooling as shown in the left panel of Fig.~\ref{fig1}. However, in the presence of viscosity, additional heat production slows down the cooling process. Similarly, the vortical motion present in the QGP medium imposes constraints on its cooling. During the initial moments of evolution, the rotation speed is nearly equal to the evolution rate of the medium. It therefore leaves cooling almost unaffected, as
illustrated in the green line in Fig.~\ref{fig1}. The cooling rate of $(T^{\omega}_{\text{SO}} )$ remains approximately equal to $(T_{\text{Ideal}})$ until \( \tau \simeq 2 \) fm. After this point, the system tries to slow the evolution process as the rotation speed falls below the fluid velocity. The combined effect of vorticity and viscosity ($T^{\omega}_{\text{SO}-\Phi}$) is shown in the blue line, which predicts a slightly faster cooling than $( T_{\text{SO}} )$ due to the
resistance of viscosity to changes in vorticity direction. Positive
initial vorticity results in faster cooling, almost following
the ideal rate. However, the presence of viscosity in the vortical
fluid slows down $T^{\omega}_{\text{SO}-\Phi}$ cooling compared to $(T_{\text{Ideal}})$. Furthermore, the effect of a time-varying magnetic field on temperature evolution is shown in the magenta line. It is almost equal to $T^{\omega}_{\text{SO}-\Phi}$. While a constant magnetic field adds more restriction on cooling and freeze-out at almost nearly the same time as $T_{\text{SO}}$. \\

The right panel of Fig.~\ref{fig1} illustrates the diffusion of vorticity for all aforementioned cases. In general, the vorticity changes sign from positive to negative. The negative value of $\omega$ in the figure indicates a change in the direction of rotation. This change occurs due to the initial rapid expansion of the medium and the restrictions imposed by the rotational motion. Essentially, the evolution of the medium induces rotation that opposes the initial vorticity. As time progresses, the vorticity also grows and diffuses in the opposite direction, eventually saturating once the medium's evolution becomes static.
Additionally, the results presented in the right panel of Fig.~\ref{fig1} suggest that cooling becomes nearly independent of vorticity when the fluid is rotating close to the speed of light. Similar to the temperature evolution, the vorticity evolution shows a very marginal difference between the rotating viscous case and the rotating viscous case with a time-varying magnetic field. However, during the initial time of medium evolution, a sudden drop of the vorticity is found for a rotating non-viscous medium and a rotating viscous medium, but with a constant magnetic field. The evolution of hydrodynamic fields, such as temperature, vorticity, viscosity, and the magnetic field, at the decoupling freeze-out hypersurface will affect experimentally measured observables, including hadron spin polarization. 

\begin{figure}[ht!]
\centering
\includegraphics[scale = 0.5]{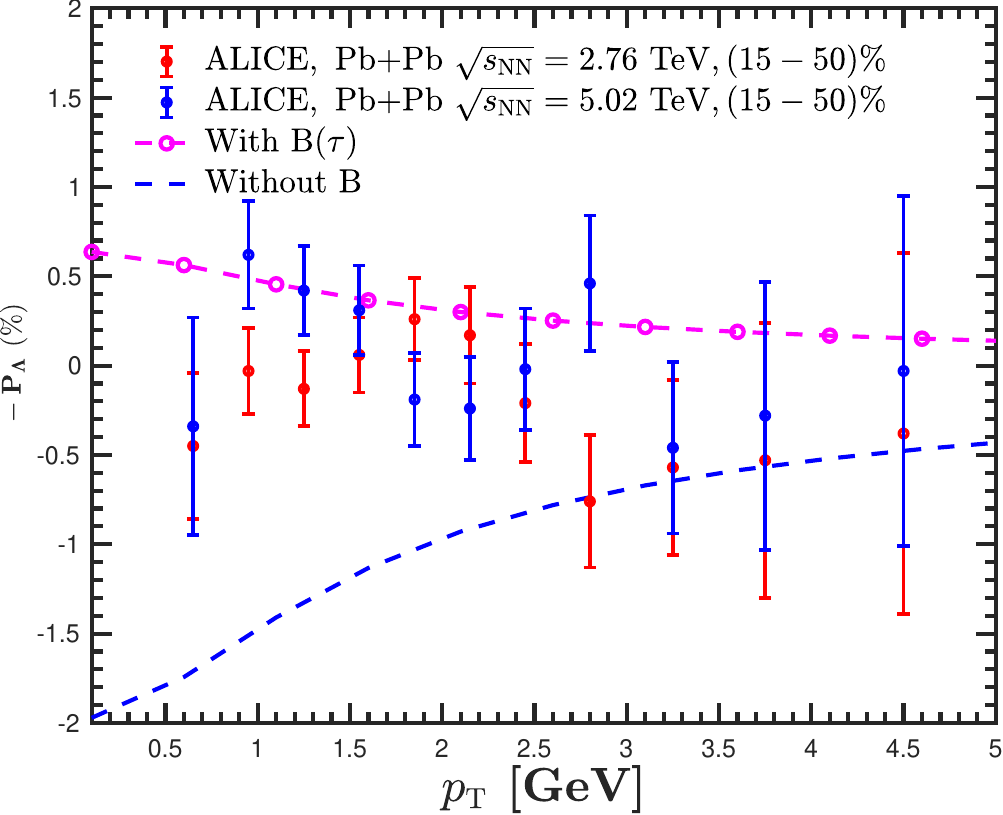}
\caption{(Color Online) The global spin polarization of $\Lambda$ hyperons as a function of transverse momentum ($p_{\rm T}$) for a rotating viscous medium with a time-varying magnetic field (magenta), and without magnetic field (blue). Our results are compared with the ALICE experimental measurement in Pb+Pb collisions at $\sqrt{s_{\rm NN}}$ = 2.76 and 5.02 TeV at (15-50)\% collision centrality~\cite{ALICE:2019onw}.}
\label{fig2}
\end{figure}

Figure~\ref{fig2} shows the global spin polarization of $\Lambda$ hyperons as a function of transverse momentum ($p_{\rm T}$) obtained using a coupled hydrodynamic framework as discussed in Sec.~\ref{formulation}. Unlike the usual hydrodynamic framework, where the vorticity profile is obtained from the solution of Israel-Stewart hydrodynamics equations, here, we couple the vorticity in the hydrodynamic equations and in the velocity profile as discussed in Sec.~\ref{formulation}. 
We derive a dissipation-rate equation for vorticity in the presence of a magnetic field, which governs the spin polarization of $\Lambda$ hyperons in two scenarios. First, for a rotating viscous medium with a time-varying magnetic field shown in the magenta line of Fig.~\ref{fig2}. Second, for a rotating viscous medium without a magnetic field, shown in blue line in Fig.~\ref{fig2}. We compare our results with the recent experimental measurement of the ALICE Collaboration in Pb+Pb collisions at $\sqrt{s_{\rm NN}}$ = 2.76 and 5.02 TeV at (15-50)\% collision centrality~\cite{ALICE:2019onw}.  The initial conditions used to solve the hydrodynamic equations are according to the compared collision system, energies, and centrality. We find a qualitative agreement between the hydrodynamic theory and experimental data. The range of spin polarization between the two cases in this study can be regarded as a measure of the theoretical uncertainty associated with different initial conditions. Our study presents a unique method to estimate the vorticity evolution and spin polarization of $\Lambda$ hyperons, offering deeper insight into the vorticity and spin polarization phenomena in relativistic heavy-ion collisions.

\section{Summary}
\label{sum}
In this work, a unified relativistic hydrodynamic framework has been developed to investigate the global spin polarization of $\Lambda$ hyperons in ultrarelativistic heavy-ion collisions. The formulation consistently incorporates medium vorticity, shear viscosity, and magnetic field effects within a second-order M\"{u}ller–Israel–Stewart (MIS) theory. It is an essential element of the present approach in the dynamical treatment of vorticity through a dissipation-rate equation, enabling a self-consistent evolution of thermal vorticity alongside temperature and viscosity. \\

The coupled non-linear evolution equations are solved using initial conditions for Pb–Pb collisions at $\sqrt{s_{\rm NN}} = 2.76$ and $5.02$ TeV. Consequently, the temperature cooling and vorticity dissipation profiles demonstrate that viscosity moderates the expansion dynamics, while vorticity introduces a non-trivial interplay with the medium evolution, particularly at later times. The role of magnetic fields is found to be marginal for time-dependent configurations, whereas a constant magnetic field introduces a somewhat stronger constraint on the system evolution. The thermal vorticity evaluated at the isothermal freeze-out hypersurface is subsequently used to compute the spin polarization of $\Lambda$ hyperons. The framework naturally separates the contributions arising from vorticity and magnetic fields. A comparison with ALICE measurements shows qualitative agreement, indicating that the coupled evolution of vorticity and hydrodynamic fields captures the essential physics governing global polarization at TeV energies. The spread between different magnetic field scenarios provides an estimate of the theoretical uncertainties associated with medium dynamics and initial conditions.\\

Altogether, the present study establishes a self-consistent connection between macroscopic fluid properties and microscopic spin observables, offering a coherent description of polarization phenomena across RHIC and LHC energies. Such developments are essential for establishing spin polarization as a precision probe of the vortical and electromagnetic structure of hot QCD matter.\\

{\it Future Outlook:} Several directions can further advance the present framework:
\begin{itemize}
\item The current implementation is based on the 2+1D expansion of the system. Extending the framework to 3+1D event-by-event viscous hydrodynamics would allow for a more realistic description of longitudinal dynamics, fluctuations, and initial state geometry.

\item  The present study uses an equation of state (EoS) describing massless quarks and gluons. However, a lattice QCD-based EoS could be more appropriate for a more comprehensive study. 

 \item The considered velocity profile provides a controlled description of rotation; however, realistic heavy-ion collisions exhibit complex flow structures. Incorporating fluctuating and non-linear velocity fields can improve the quantitative description of vorticity generation.

 \item The magnetic field is modeled through simplified profiles. A fully dynamical magnetohydrodynamic treatment, including conductivity and medium response, would provide a more accurate estimate of electromagnetic effects on polarization.

 \item The present approach treats spin polarization as a probe for QGP. Extending the framework to include spin as an explicit hydrodynamic degree of freedom (spin hydrodynamics) would enable the study of backreaction effects and spin transport phenomena.

 \item The polarization information can be modified during the hadronic stage through rescattering and resonance decays. Incorporating hadronic afterburners and feed-down effects is essential for direct comparison with experimental observables.

 \item The strong dependence of initial conditions on medium evolution motivates a systematic exploration using Bayesian inference techniques to constrain model parameters such as thermalization time, medium initial viscosity, initial vorticity, and magnetic field, etc.,  with experimental data.

 \item The framework can be extended to study local polarization, vector-meson spin alignment, correlations, and fluctuations of spin polarization, providing a more comprehensive understanding of spin dynamics in QCD matter.
\end{itemize}

\section*{Acknowledgement}
 The authors acknowledge Jan-e Alam for fruitful discussion during the preparation of the manuscript. Bhagyarathi Sahoo acknowledges the financial aid from CSIR, Government of India. The authors gratefully acknowledge the DAE-DST, Government of India, funding under the mega-science project "Indian Participation in the ALICE experiment at CERN" bearing Project No. SR/MF/PS-02/2021-IITI (E-37123).

\appendix
\section{Calculation of $D \equiv u^{\mu}\partial_{\mu}$}
\label{D}
The convective time derivative $u^{\mu}\partial_{\mu}$ is obtained as;

\begin{equation}   
D \equiv u^{\mu}\partial_{\mu} = u^{0}\partial_{0} + u^{x}\partial_{x} + u^{y}\partial_{y} + u^{z}\partial_{z}
\end{equation}

Using the four velocity $u^{\mu} =\gamma(1, v_{x}, 0, v_{z}) = \gamma \bigg(1, \frac{\omega z}{2}, 0, \frac{z}{\tau} - \frac{\omega x}{2} \bigg)$, we have

\begin{equation}   
u^{\mu}\partial_{\mu} = \gamma \bigg( \frac{\partial}{\partial t} + v_{x} \frac{\partial}{\partial x}+ v_{z} \frac{\partial}{\partial z} \bigg)
\label{umudelmu}
\end{equation}

\begin{equation}   
u^{\mu}\partial_{\mu} = \gamma \bigg[ \frac{\partial}{\partial t} +\frac{\omega z}{2} \frac{\partial}{\partial x}+ \bigg(\frac{z}{\tau} - \frac{\omega x}{2}\bigg) \frac{\partial}{\partial z} \bigg]
\label{umudelmu}
\end{equation}

where $x$ and $z$ are the position coordinates. Here, it is important to note that vorticity is the cause of inducing the velocity along the $x$-direction, i.e., $v_{x}$.\\

\section{Calculation of $\theta \equiv \partial_{\mu}u^{\mu}$}
\label{theta}
The expansion rate $\partial_{\mu}u^{\mu}$ explore the velocity grandient, as given by;
\begin{equation}   
\theta \equiv \partial_{\mu}u^{\mu} = \partial_{0}u^{0} + \partial_{x}u^{x} + \partial_{y}u^{y} + \partial_{z}u^{z}
\end{equation}

With the help of four velocity $u^{\mu}$, we get;

\begin{equation}   
\partial_{\mu}u^{\mu} = \bigg( \frac{\partial \gamma}{\partial t} + \frac{\partial  ( \gamma v_{x} )}{\partial x}+ \frac{\partial (\gamma v_{z})}{\partial z} \bigg)
\label{delmuumu}
\end{equation}
Making a coordinate transformation from cartesian to Milne coordinate, we have
$(t,x,y,z) \rightarrow (\tau, r, \phi, \eta)$.
In Milne Coordinate system $t = \tau \cosh\eta$, $z = \tau \sinh\eta$, 

\begin{equation}
\frac{\partial }{\partial t} = \cosh \eta \frac{\partial }{\partial \tau} - \frac{\sinh \eta}{\tau} \frac{\partial }{\partial \eta}
\label{deldelt}
\end{equation}

\begin{equation}
\frac{\partial }{\partial z} = - \sinh \eta \frac{\partial }{\partial \tau} + \frac{\cosh \eta}{\tau} \frac{\partial }{\partial \eta}
\label{deldelz}
\end{equation}

Solving the volume expansion term ($\theta$) using the velocity profile mentioned in Eq.~\ref{vxz} in Milne coordinate, we obtain; 
\begin{widetext}

\begin{align}
\partial_{\mu}u^{\mu} = \bigg[  \frac {\gamma \cosh^2 \eta}{\tau} + \frac{\gamma^{3} \omega^2 \tau^2 \sinh^2 \eta}{8} \left(\frac{\omega^2 x}{2}  - \omega \sinh \eta  \right) + & \frac{\gamma^3}{2 \tau} \left( (\sinh \eta - \frac{\omega x}{2}) \cosh \eta - \sinh \eta \right)\left( (\frac{\omega^2 \tau^2}{4} +1 ) \sinh 2\eta - \omega x\right) \nonumber \\ + \frac{\gamma^3 \omega^2 \tau \sinh^2 \eta }{4}\left( \cosh \eta - (\sinh \eta - \frac{\omega x}{2})\sinh \eta \right) \bigg]
\label{delmuumu2}  
\end{align}   

\section{Temperature cooling}
\label{tempcooling}

Substituting the expression of $D$ in Eq.~\ref{DepsilonB}, we have
\begin{align}
& \left( \cosh \eta - \sinh^{2} \eta + \frac{ \omega x \sinh \eta}{2} \right) \frac{\partial \epsilon }{\partial  \tau} + \frac{1}{\tau}\left( (\cosh \eta - 1) \sinh \eta - \frac{ \omega x \cosh \eta}{2} \right) \frac{\partial \epsilon }{\partial  \eta} +  \frac{ \omega \tau \sinh \eta}{2} \frac{\partial \epsilon }{\partial x} \nonumber \\
& =  - \frac{1}{\gamma} \left(\epsilon+P+B^2 - \Phi \right) \times (\partial_{\mu} u^{\mu})  - B \bigg[ \left( \cosh \eta - \sinh^{2} \eta + \frac{ \omega x \sinh \eta}{2} \right) \frac{\partial B }{\partial  \tau} \nonumber \\
& -  \frac{1}{\tau}\left( (\cosh \eta - 1) \sinh \eta - \frac{ \omega x \cosh \eta}{2} \right) \frac{\partial B}{\partial  \eta} -  \frac{ \omega \tau \sinh \eta}{2} \frac{\partial B }{\partial x} \bigg]
\label{energyevolution}
\end{align}

To obtain the temperature cooling from Eq.~\ref{energyevolution}, we use an equation of state (EoS) such that pressure $P = \epsilon/3= aT^{4}$, and we obtain 
\begin{align}
& \left( \cosh \eta - \sinh^{2} \eta + \frac{ \omega x \sinh \eta}{2} \right)  \frac{\partial T }{\partial  \tau} + \frac{1}{\tau}\left( (\cosh \eta - 1) \sinh \eta - \frac{ \omega x \cosh \eta}{2} \right) \frac{\partial T }{\partial  \eta} +  \frac{ \omega \tau \sinh \eta}{2} \frac{\partial T }{\partial x} \nonumber \\
& =  - \frac{1}{\gamma} \left[ \frac{T}{3} \bigg( 1 + \frac{2\omega T^{2}}{s\pi^{2}}\cosh\left(\frac{\omega}{2T}\right)   + \frac{\chi_{m}eB^{2}}{Ts} \bigg) +\frac{\Phi T^{-3}}{12a} \right] (\partial_{\mu} u^{\mu}) 
- \frac{B T^{-3}}{12a} \bigg[ \left( \cosh \eta - \sinh^{2} \eta + \frac{ \omega x \sinh \eta}{2} \right) \times \nonumber \\
& \frac{\partial B }{\partial  \tau} -  \frac{1}{\tau}\left( (\cosh \eta - 1) \sinh \eta - \frac{ \omega x \cosh \eta}{2} \right) \frac{\partial B}{\partial  \eta} -  \frac{ \omega \tau \sinh \eta}{2} \frac{\partial B }{\partial x} \bigg]
\label{tempevolution}
\end{align}

\section{Shear viscous term evolution}
\label{shearviscous}
Similarly, the evolution of shear viscous term ($\Phi = \pi^{00} - \pi^{zz}$) can be obtained from the  Eq.~\ref{phievolution},

\begin{align}
&\left( \cosh \eta - \sinh^{2} \eta + \frac{ \omega x \sinh \eta}{2} \right) \frac{\partial \Phi }{\partial  \tau} + \left( (\cosh \eta - 1) \sinh \eta - \frac{ \omega x \cosh \eta}{2\tau} \right) \frac{\partial \Phi}{\partial  \eta} +  \frac{ \omega \tau \sinh \eta}{2} \frac{\partial \Phi}{\partial x} \nonumber\\
&= - \frac{\Phi}{2 \gamma} (\partial_{\mu} u^{\mu})  - \frac{\Phi}{\gamma\tau_{\pi}}  + \frac{5 \Phi}{ 2T} \bigg[  \left( \cosh \eta - \sinh^{2} \eta + \frac{ \omega x \sinh \eta}{2} \right)  \frac{d T}{d\tau}  \nonumber \\
& + \left( (\cosh \eta - 1) \sinh \eta - \frac{ \omega x \cosh \eta}{2\tau} \right) \frac{\partial T}{\partial  \eta}
+ \frac{ \omega \tau \sinh \eta}{2} \frac{\partial T}{\partial x}  \bigg] +  \frac{1}{\gamma\beta_{2}}\left(\bigtriangledown^{<0}u^{0>} - \bigtriangledown^{<z}u^{z>}\right)
\label{phievolution2}
\end{align}

The form of  $\bigtriangledown^{<0}u^{0>} - \bigtriangledown^{<z}u^{z>}$ can be found in Ref.~\cite{Sahoo:2023xnu}. 

\section{Vorticity dissipation rate}
\label{vordiss}

The vorticity dissipation rate can be obtained from Eq.~\ref{wevolution} as follows
\begin{align}
\label{wevolution2eB}
& \left( \cosh \eta - \sinh^{2} \eta + \frac{ \omega x \sinh \eta}{2} \right) \frac{ \partial \omega}{\partial \tau} + \left( (\cosh \eta - 1) \sinh \eta - \frac{ \omega x \cosh \eta}{2\tau} \right) \frac{\partial \omega}{\partial  \eta} +  \frac{ \omega \tau \sinh \eta}{2} \frac{\partial \omega}{\partial x} \nonumber\\
&= \frac{-\pi^{2}}{2G} \bigg[ \frac{4T}{3 \gamma }\bigg( s + \frac{2T^{2}w}{\pi^{2}} \rm cosh\left(\frac{\omega}{2T}\right) + (1 + e\chi_m)\frac{B^2}{T} - \frac{\Phi}{T}\bigg) \times (\partial_{\mu} u^{\mu}) + \bigg(s+3dT^{3} + \frac{2F}{\pi^{2}}\bigg)  \times \{ \left( \cosh \eta - \sinh^{2} \eta + \frac{ \omega x \sinh \eta}{2} \right) \frac{ \partial T}{\partial \tau} \nonumber \\
& + \left( (\cosh \eta - 1) \sinh \eta - \frac{ \omega x \cosh \eta}{2\tau} \right) \frac{\partial T}{\partial  \eta} +  \frac{ \omega \tau \sinh \eta}{2} \frac{\partial T}{\partial x} \} + \bigg(\frac{4}{3} + 2e \chi_m \bigg) B \{ \left( \cosh \eta - \sinh^{2} \eta + \frac{ \omega x \sinh \eta}{2} \right) \frac{ \partial B}{\partial \tau} \nonumber \\
& + \left( (\cosh \eta - 1) \sinh \eta - \frac{ \omega x \cosh \eta}{2\tau} \right) \frac{\partial B}{\partial  \eta} +  \frac{ \omega \tau \sinh \eta}{2} \frac{\partial B}{\partial x} \} \bigg]
\end{align}
\section{Freeze-out hypersurface}
\label{hypersurface}

Now, we will discuss the details of the decoupling hypersurface element $d\Sigma_{\lambda}$ in cylindrical symmetry, which can be written as~\cite{Ruuskanen:1986py},

\begin{equation}
d\Sigma^{\mu} = (rd\phi dr dz, \hat{e}_r rd\phi dz dt, 0, \hat{e}_z rd\phi dr dt)
\end{equation}
with $u^{\mu}$ can be written as, 
\begin{equation}
u^{\mu} = \cosh y_T(\cosh \eta, \tanh y_T, 0, \sinh \eta)
\label{fourvelocity}
\end{equation}
where $y_T$ is the transverse rapidity.
We replace the space-time rapidity $\eta$ with the longitudinal flow rapidity $\theta$, and we use $p_T$ to fix the direction of the x-axis in the transverse plane. Then 
\begin{equation}
p_{\mu}u^{\mu} = m_T \cosh y_T \cosh (\theta - y) - p_T \sinh y_T\cos \phi
\label{pmuumu}
\end{equation}
and

\begin{align}
p_{\mu}d\Sigma^{\mu} =rd\phi d\eta \left( m_T \tau \cosh (\eta -y)) dr +  m_T \sinh (\eta -y) d\tau\right)
- r\;p_T\; \tau\; cos\phi\; d\phi \; d\eta\; d\tau 
\label{pmudsigmamu}
\end{align}
\vspace{0.2 cm}

where $m_{T}$ = $\sqrt{ m^2 + p_{\rm T}^2 }$ is the transverse mass, and y is the rapidity of a particle of momentum $p^{\mu}$. For the boost invariant flow $\theta$ = $\eta$, that leads to the second of Eq.~\ref{pmudsigmamu} vanishes.
The distribution function $f(x,p)$ is modified using $p_{\mu}u^{\mu}$, given as;
\begin{equation}
f(x,p) = \frac{1}{\exp{\left[\left( m_T \cosh y_T \cosh ( \eta - y) - p_T \sinh y_T\cos \phi \right)/T \right]} +1}
\end{equation}
Substituting the expression of $p_{\mu}d\Sigma^{\mu}$ and  $p_{\mu}u^{\mu}$ in Eq.~\ref{spinpolarization3} at zero chemical potential, the average mean-spin vector can be obtained as: \\
\begin{equation}
S_y  = \frac{1}{4} \frac{\displaystyle\int \tau r \displaystyle\int_{0}^{2\pi} d\phi \displaystyle\int_{-\infty}^{+\infty} d\eta\; \;\bigg[\bar{\omega}_{y} - \frac{2 \mu_{\Lambda}  B_{y}}{T} \bigg] f(x,p)\left( 1-  f(x,p) \right)(m_T \cosh (\eta -y) dr - p_T\;cos\phi\;d\tau)} {\displaystyle\int \tau r \displaystyle\int_{0}^{2\pi} d\phi \displaystyle\int_{-\infty}^{+\infty} d\eta \;f(x,p)\; (m_T \cosh (\eta -y) dr - p_T\;cos\phi\;d\tau)\; dr}
\label{spinpolarization5}
\end{equation}

Since the mass of the $\Lambda$ is much larger than the
temperature range being considered in this study, we assume that 1 - f(x,p) $\approx 1 $. The integration over $\tau$ and $\phi$ can be carried out if we expand the distribution function in a geometric series. On simplification, we have
\begin{equation}
S_y  = \frac{1}{4} \frac{\displaystyle\int \tau r \displaystyle\sum_{n=1}^{\infty} E\;\bigg[\bar{\omega}_{y} - \frac{2 \mu_{\Lambda}  B_{y}}{T} \bigg] \left[ m_T I_0 \left(O_1\right)  K_1 \left(O_2 \right) dr - p_T I_1 \left(O_1 \right)  K_0 \left(O_2 \right) d\tau \right] }{\displaystyle\int \tau r \displaystyle\sum_{n=1}^{\infty} \left[ m_T I_0 \left(O_1 \right)  K_1 \left(O_2 \right) dr - p_T I_1 \left(O_1 \right)  K_0 \left(O_2 \right) d\tau \right]} 
\label{spinpolarization6}
\end{equation}

where I's, K's are Bessel functions and $O_{1} = n \frac{p_T}{T} \sinh y_T$, $O_{2} = n \frac{m_T}{T} \cosh y_T$.
We assume the hadronization happens in the decoupling hypersurface where $\tau$ = constant, $T = T_{dec}$\\
\begin{equation}
S_y  = \frac{1}{4 \; T_{\rm dec}} \frac{\displaystyle\int   \tau r\; \displaystyle\sum_{n=1}^{\infty} \; (\omega_{y} - 2 \mu_{\Lambda} B_{y})  m_T I_0 \left( O_{1}\right)  K_1 \left( O_{2}\right) dr }{\displaystyle\int \tau r  \displaystyle\sum_{n=1}^{\infty}  m_T I_0 \left(O_1 \right)  K_1 \left(O_2 \right) dr}
\label{spinpolarization8}
\end{equation}

\end{widetext}

\end{document}